\documentclass[12pt,twoside,a4paper]{article}
\usepackage{amsmath,amssymb,latexsym,theorem,bbm,shapepar,natbib,epsfig,color}
\newfont{\msa}{msam10 scaled\magstep1}
\newfont{\ssmsa}{msam9}

\def\DS{\mathop{\hbox{\rm DS}}}
\def\ES{\mathop{\hbox{\rm ES}}}
\def\EE{\mathop{\hbox{\rm EE}}}

\numberwithin{equation}{section}

\title{Joint probabilistic forecasting of wind speed and temperature
  using Bayesian model averaging}

\author{{\sc S\'andor Baran$^{1}$} and {\sc Annette M\"oller$^{2}$}\\
         $^1$Faculty of Informatics, University of Debrecen\\
         Kassai \'ut 26, H-4028 Debrecen, Hungary \\
         $^2$ Department of Animal Sciences, University of
         G\"ottingen\\
         Carl-Sprengel-Weg 1, D-37075 G\"ottingen, Germany}

\date{}

\begin{document}
\pagestyle{myheadings}

\maketitle

\begin{abstract}

Ensembles of forecasts are typically employed to account for the
forecast uncertainties inherent in predictions of future weather states.
However, biases and dispersion errors often present in forecast
ensembles require statistical post-processing.
Univariate post-processing models such as Bayesian model averaging
(BMA) have been successfully applied for various weather quantities.
Nonetheless, BMA and many other standard post-processing procedures are
designed for a single weather variable, thus ignoring possible
dependencies among weather quantities.
In line with recently upcoming research to develop multivariate
post-processing procedures, e.g., BMA for bivariate wind vectors, or
flexible procedures applicable for multiple weather quantities of
different types, a bivariate BMA model for joint calibration of wind
speed and temperature forecasts is proposed based on the
bivariate truncated normal distribution. It extends the univariate
truncated normal BMA model designed for post-processing ensemble forecast
of wind speed by adding a normally distributed temperature component with a
covariance structure representing the dependency among the two weather
quantities.

The method is applied to wind speed and temperature forecasts of the
eight-member University of Washington mesoscale ensemble and of the
eleven-member ALADIN-HUNEPS ensemble of the Hungarian Meteorological
Service and  its predictive performance is compared to that of the general
Gaussian copula method. The results indicate improved calibration of
probabilistic and accuracy of point forecasts in comparison to the raw
ensemble and the overall 
performance of this model is able to keep up with that of the Gaussian
copula method.

\end{abstract}

\section{Introduction}
   \label{sec:sec1}
The main objective of weather forecasting is to give a
reliable prediction of future atmospheric states on the basis of
observational data, prior forecasts valid for the initial time
of the forecasts and mathematical models describing the dynamical and
physical behaviour of the atmosphere.
These models numerically solve the set of the hydro-thermodynamic
non-linear partial differential equations of the atmosphere and its
coupled systems. A disadvantage of
these numerical weather prediction models is that since the
atmosphere has a chaotic character the solutions
strongly depend on the initial conditions and also on other uncertainties
related to the numerical weather prediction process. In practice it
means that the results of such models are never fully accurate and the
forecast uncertainties should be also taken into account in the forecast
preparation.
One can reduce the uncertainties by running the model with different
initial conditions and produce an
ensemble of forecasts. Using a forecast ensemble one can estimate the
probability distribution of future weather variables which allows
probabilistic weather forecasting \citep{gr}, where not only the
future atmospheric states are
predicted, but also the related uncertainty information such as
variance, probabilities of various events, etc.
The ensemble
prediction method was proposed by \citet{leith} and since its first
operational implementation \citep{btmp,tk} it became a widely used
technique all over the world \citep[see,
e.g.,][]{em05,lp,gtpb,horanyi}. However, although, e.g., the ensemble mean
on average yields better forecasts of a meteorological quantity than
any of the individual ensemble members, it is often the case that the
ensemble is under-dispersive
and in this way, uncalibrated \citep{bhtp}, therefore calibration is
absolutely needed to account for this deficiency.

The Bayesian model averaging (BMA) method for calibrating
forecast ensembles was introduced by \citet{rgbp}. The BMA predictive
probability density function (PDF) of a future weather quantity is the
weighted sum of individual PDFs corresponding to the ensemble
members. An individual PDF can be interpreted as the
conditional PDF of the future weather quantity provided the considered
forecast is the best one and the weights are based on the
relative performance of the ensemble members during a given training
period. In this way BMA is a special, fixed parameter version of
dynamic model averaging method developed by \citet{rke}. Weights
and other model parameters are usually estimated using linear
regression and maximum likelihood (ML)
method, where the maximum of the likelihood function is mostly found by EM
algorithm.  In practice, the performance of the individual
ensemble members should have a clear characteristic (and not a random
one) or if it is not the case (the ensemble members are exchangeable)
this fact should be taken into account
at the calibration process \citep[see, e.g.,][]{frg}. In \citet{rgbp} the BMA
method was successfully applied to obtain 48-hour forecasts of surface
temperature and sea level pressure in the North American Pacific
Northwest based on the 5 members of the University of Washington
mesoscale ensemble \citep{gm}. These weather quantities can be
modeled by normal distributions, so the predictive PDF is a Gaussian mixture.
Later, \citet{srgf} developed a discrete-continuous BMA model for
precipitation
forecasting, where the discrete part corresponds to the event of no
precipitation, while the cubic root of the precipitation amount
(if it is positive) is modeled by a gamma distribution. In
\citet{sgr10} the BMA method was used for wind speed forecasting and
the component PDFs follow generalized gamma distributions. Using a von Mises
distribution to model angular data,
\citet{bgrgg} introduced a BMA scheme to predict surface wind
direction. Finally, \citet{bar} suggests the use of truncated normal
mixture for modelling wind speed.

Another possible method for statistical post-processing of ensemble forecasts
is the ensemble model output statistics (EMOS) introduced by
\citet{grwg} for calibrating forecasts of weather quantities following a
normal distribution (sea level pressure, temperature). For these
weather variables the EMOS model produces a single normal PDF, where the
mean and the variance depend on the ensemble members. Later, \citet{tg}
extended this method to truncated normal distributions and
used it for calibrating wind speed data, while \citet{sch} developed
an EMOS model for precipitation forecasting.

All models mentioned above consider only a single weather quantity and
recently an increasing interest has appeared in investigating the
correlation between different variables. For calibrating wind vector
forecasts, which can be modeled using a bivariate normal distribution,
\citet{pinson} suggested an adaptive technique, \citet{sgr13}
described a BMA model, while \citet{stg} developed an EMOS approach.
A different idea appears in \citet{mlt}, where after performing
separate univariate calibrations the authors use a Gaussian copula to
preserve the dependence between the weather variables
investigated. Finally, for exchangeable ensembles \citet{stg13}
introduced the ensemble copula coupling method (ECC) which after univariate
calibration uses the rank order information available in the raw
ensemble.

In the present paper we develop a BMA model for joint
calibration of ensemble forecasts of wind speed and temperature. In
our approach the predictive PDF is a mixture of bivariate normal distributions
truncated in the first (wind) coordinate from below at the origin. For
parameter estimation we use the ML method and the likelihood function
is maximized with the help of truncated data EM algorithm for Gaussian
mixture models \citep{ls12}.

We test our model on the ensemble forecasts of maximum wind speed and
daily minimum temperature produced by the eight-member University of
Washington mesoscale ensemble \citep[UWME;][]{em05}, and compare the
results with the performance of the Gaussian copula method suggested
by \citet{mlt}. Additionally, we perform tests with the
operational Limited
Area Model Ensemble Prediction System of the Hungarian
Meteorological Service (HMS) called ALADIN-HUNEPS \citep{hagel,
  horanyi}. We remark that similarly to the UWME, univariate BMA
calibration of wind speed \citep{bhn1,bar} and temperature
\citep{bhn2} forecasts of the ALADIN-HUNEPS system have already been
investigated.

\section{Data}
  \label{sec:sec2}

\subsection{University of Washington mesoscale ensemble}
  \label{subs:subs2.1}
The eight-member University of Washington mesoscale ensemble covers the Pacific
Northwest region of western North America providing forecasts on a 12
km grid. The ensemble members are
obtained from different runs of the fifth generation Pennsylvania State
University--National Center for Atmospheric Research mesoscale model
(PSU-NCAR MM5) with initial conditions from different sources
\citep{grell}. Our data base (identical to the one used in
\citet{mlt}) contains ensembles of 48-hour forecasts
and corresponding validation observations of 10 meter maximum wind
speed (maximum of the hourly instantaneous wind speeds over the
previous twelve hours, given in m/s, see, e.g., \citet{sgr10}) and 2
meter minimum temperature (given in K) for 152 stations in the Automated Surface
Observing Network \citep{asos} in the US states of Washington, Oregon,
Idaho, California and Nevada for calendar years 2007 and 2008. The
forecasts are
initialized at 0 UTC (5 pm local time when DST is in use
and 4 pm otherwise) and the generation of the ensemble ensures that
its members are not exchangeable. In the present study we investigate
only forecasts for calendar year 2008 with additional data from the
last two months of 2007 used for parameter estimation. After removing
days and locations with missing data, 92 stations remained where
the number of days for which forecasts and validating observations are
available varies between 141 and 290.

\subsection{ALADIN-HUNEPS ensemble}
  \label{subs:subs2.2}
The ALADIN-HUNEPS system of the HMS covers a large part of Continental
Europe with a
horizontal resolution
of 12 km and it is obtained by dynamical downscaling (by the ALADIN
limited area model) of the global
ARPEGE based PEARP system of M\'et\'eo France \citep{hkkr,dljn}. The
ensemble consists of 11 members, 10 initialized from perturbed initial
conditions and one control member from the unperturbed analysis,
implying that the ensemble contains groups of exchangeable
forecasts. The data base contains 11 member ensembles of 42-hour
forecasts for 10 meter instantaneous wind speed (given
in m/s) and 2 meter temperature (given in K) for 10 major cities in
Hungary (Miskolc, Szombathely, Gy\H or, Budapest, Debrecen, Ny\'\i regyh\'aza,
Nagykanizsa, P\'ecs, Kecskem\'et, Szeged) produced by the
ALADIN-HUNEPS system of the HMS, together with the corresponding
validating observations for the one-year period between April 1, 2012
and March 31, 2013. The forecasts are
initialized at 18 UTC (8 pm local time when daylight saving time (DST)
operates and 7 pm otherwise). The data set is fairly complete since there are
only six days when no forecasts are available and these days have been
excluded from the analysis.

\section{Methods}
  \label{sec:sec3}
\subsection{Bayesian model averaging}
  \label{subs:subs3.1}

Denote by \ $f_1,f_2,\ldots ,f_M$ \ the ensemble forecast of a
certain weather quantity (vector)\ $X$ \ for a given location and time.
In the BMA model for ensemble forecasting
\citep{rgbp}, to each ensemble member \
$f_k$ \ corresponds a component PDF \ $g_k(x \vert f_k, \theta_k)$, \
where \ $\theta_k$ \ is a parameter to be estimated. The BMA
predictive PDF of \ $X$ \ is
\begin{equation}
  \label{eq:eq3.1}
p(x\vert\, f_1, \ldots ,f_M;\theta_1, \ldots
,\theta_M):=\sum_{k=1}^M\omega _k g_k(x \vert\, f_k, \theta_k),
\end{equation}
where the weight \ $\omega_k$ \ is connected to the relative
performance of the ensemble member \ $f_k$ \ during the training
period. Obviously, these weights form a probability distribution, that
is \ $\omega_k\geq 0, \ k=1,2, \ldots ,M,$ and $\sum_{k=1}^M
\omega_k=1$. \

Once the predictive density \eqref{eq:eq3.1} is given,
one can take its mean or median as a point forecast for \ $X$. \ We
remark, that for a $d$-dimensional distribution function \ $F$ \ a
multivariate median  is a vector minimizing the function
\begin{equation*}
\phi(\boldsymbol\alpha):=\int_{\mathbb R^d} \Vert \boldsymbol\alpha -\boldsymbol
x\Vert F({\mathrm d}\boldsymbol x),
\end{equation*}
where \ $\Vert\cdot\Vert$ \ denotes the Euclidean norm, and if $F$ is
not concentrated on a line in $\mathbb R^d$ then the median is
unique \citep{md}.

BMA model \eqref{eq:eq3.1} is valid only in the cases when the sources
of the ensemble members are clearly distinguishable, as for the UWME
\citep{em05} or for the
COSMO-DE ensemble of the German Meteorological Service \citep{gtpb}.
However, most of the currently used ensemble
prediction systems  produce ensembles where some ensemble
members are statistically indistinguishable. Usually, these
exchangeable ensemble members are obtained with the help of
perturbations of the initial conditions, which is the case for the 51
member European Centre for Medium-Range Weather Forecasts
ensemble \citep[ECMWF;][]{lp} or for the ALADIN-HUNEPS ensemble described in
Section \ref{subs:subs2.2}.

Suppose we have \ $M$ \ ensemble members divided into \ $m$ \ exchangeable
groups, where the \ $k$th \ group contains \ $M_k\geq 1$ \ ensemble
members, so \ $\sum_{k=1}^mM_k=M$. \ Further, denote by \ $f_{k,\ell}$
\ the  $\ell$th member of the $k$th group. For this
situation \citet{frg} suggested to use model
\begin{equation}
  \label{eq:eq3.2}
p(x\vert f_{1,1}, \ldots ,f_{1,M_1}, \ldots ,  f_{m,1}, \ldots
,f_{m,M_m} ;\theta_1, \ldots
,\theta_m):=\sum_{k=1}^m\sum_{\ell=1}^{M_k}\omega _k g_k(x \vert\,
f_{k,\ell}, \theta_k),
\end{equation}
where ensemble members within a given group have the same weights
and parameters.

To simplify notations we give the results and
formulae of this section for model \eqref{eq:eq3.1}, but their
generalization to model
\eqref{eq:eq3.2} is rather straightforward.

\subsection{Bivariate truncated normal model}
  \label{subs:subs3.2}

As it has already been mentioned in the Introduction, for temperature
observations a BMA model with normal component PDFs can be fit
reasonably well, while for wind speed observations BMA methods with
gamma \citep{sgr10} and with truncated normal components \citep{bar},
respectively, had been developed. This gives the natural idea of joint
modelling wind speed and temperature with a bivariate normal
distribution with first (wind) coordinate truncated from below at
zero. If
\begin{equation*}
\boldsymbol\mu=\begin{bmatrix}\mu_W \\ \mu_T \end{bmatrix} \qquad
\text{and} \qquad \Sigma=\begin{bmatrix}\sigma^2_W& \sigma_{WT}
  \\\sigma_{WT}&\sigma^2_{T} \end{bmatrix}
\end{equation*}
are the location vector and scale matrix, respectively, provided
 \  $\Sigma$ \ is regular, the
joint PDF of this special bivariate truncated normal distribution
\ ${\mathcal N}_2^{\, 0}(\boldsymbol \mu, \Sigma)$ \ is
\begin{equation}
  \label{eq:eq3.3}
g(\boldsymbol x| \boldsymbol\mu,\Sigma )\!:=\!\frac
{\big(\det(\Sigma)\big)^{-1/2}}
{2\pi\Phi\big(\mu_W/\sigma_W\big)}\exp\Big(-\frac
12(\boldsymbol x-\boldsymbol\mu)^{\top}\Sigma^{-1}(\boldsymbol
x-\boldsymbol\mu)\Big){\mathbb I}_{\{x_W\geq 0\}}, \quad \boldsymbol
x\!=\begin{bmatrix}x_W \\x_T \end{bmatrix}\! \in\! {\mathbb R}^2,
\end{equation}
where \ $\Phi$ \ denotes the cumulative distribution function (CDF) of the
standard normal distribution and by \ ${\mathbb I}_H$ \ we denote the indicator
function of a set \ $H$. \ The mean vector \
$\boldsymbol\kappa$ \ and
covariance matrix \ $\Xi$ \ of \
\ ${\mathcal N}_2^{\, 0}(\boldsymbol \mu, \Sigma)$  \ are
\begin{equation*}
\boldsymbol\kappa\!=\!\boldsymbol\mu
+\frac{\sigma_W\varphi\big(\mu_W/\sigma_W\big)}{\Phi\big(\mu_W/\sigma_W\big)}
\begin{bmatrix}1 \\0 \end{bmatrix}
\quad \text{and} \quad \Xi\!=\!\Sigma\!-\!\left(
  \frac{\mu_W}{\sigma_W}\frac{\varphi\big(\mu_W/\sigma_W\big)}{\Phi
    \big(\mu_W/\sigma_W\big)}
  \!+\!\Bigg(\frac{\varphi\big(\mu_W/\sigma_W\big)}{\Phi
    \big(\mu_W/\sigma_W\big)}\Bigg)^2\right) \begin{bmatrix}\sigma^2_W& 0
  \\0& 0 \end{bmatrix} ,
\end{equation*}
respectively, where \ $\varphi$ \ denotes the PDF of the
standard normal distribution.

By assuming that location vector \ $\boldsymbol\mu_k$ \ of the $k$th
component PDF of the BMA
mixture \eqref{eq:eq3.1} is an affine
function of the corresponding ensemble member $\boldsymbol f_k$ and
that the scale matrices of all components are
equal, we obtain model
\begin{equation}
  \label{eq:eq3.4}
p(\boldsymbol x\vert\, \boldsymbol f_1, \ldots ,\boldsymbol f_M;A_1, \ldots
,A_M; B_1, \ldots ,B_M; \Sigma):=\sum_{k=1}^M\omega _k g(\boldsymbol x
\vert\, A_k+B_k \boldsymbol f_k,
\Sigma ),
\end{equation}
where \ $g$ \ is the PDF defined by \eqref{eq:eq3.3}, \
$A_k\in{\mathbb R}^2$ \ and \ $B_k$ \ is a two-by-two real matrix.  In this
way model
\eqref{eq:eq3.4} is a direct extension of the univariate BMA models of
temperature and wind speed investigated in \citet{rgbp} and
\citet{bar} where the authors also used the assumption of a common scale
parameter for all BMA components. It reduces the number of parameters
and makes computations easier.

One can have an even more parsimonious model by using the same
location parameters for all ensemble members, resulting in the
predictive PDF
\begin{equation}
  \label{eq:eq3.5}
q(\boldsymbol x\vert\, \boldsymbol f_1, \ldots ,\boldsymbol f_M;A; B;
\Sigma):=\sum_{k=1}^M\omega _k g(\boldsymbol x
\vert\, A+B \boldsymbol f_k,
\Sigma ).
\end{equation}
We remark that a similar type of simplification is used in the wind
speed model of the {\tt ensembleBMA} package of R \citep{frgs,frgsb}.

\subsection{Parameter estimation}
  \label{subs:subs3.3}

Model parameters \ $A_k, \ B_k, \ \omega_k, \ k=1,2,\ldots M$, \  and
\ $\Sigma$ \ of PDF \eqref{eq:eq3.4} and \ $A,\ B, \ \Sigma$ \ and \
$\omega_k, \ k=1,2,\ldots M,$ \ of PDF \eqref{eq:eq3.5} are estimated
using training data consisting of ensemble
members and validating observations from the preceding \ $n$ \ days
(rolling training period). In
what follows, \ $\boldsymbol f_{k,s,t}$ \ denotes the $k$th ensemble
member vector for location \ $s\in{\mathcal S}$ \ and  time \
$t\in{\mathcal T}$ \
and by \ $\boldsymbol x_{s,t}$ \ we denote the corresponding validating
observation.

In BMA modelling of uni- and multivariate weather quantities the location
parameters are usually estimated with linear regression of the
validating observations on the corresponding ensemble members
from the training period  \citep[see, e.g.,][]{rgbp,sgr10,sgr13}, while the
estimates of weights \ $\omega_k$ \ and scale parameter \ $\Sigma$ \
are calculated by maximizing the likelihood function of the training
data using mainly the EM algorithm for mixtures \citep{dlr,mclk}. However, this
approach assumes that the location parameter equals the mean or it can
easily be derived from it, which is not the case for the truncated normal
distribution. Hence, we suggest a pure maximum likelihood method (ML) for
estimating all parameters \citep{sgr10, bar}.

\subsubsection*{Full model}
Under the assumption of independence of forecast
errors in space and time the log-likelihood function corresponding to
model \eqref{eq:eq3.4} equals
\begin{equation}
  \label{eq:eq3.6}
\ell (\omega_1,\ldots ,\omega_M;A_1, \ldots ,A_M; B_1,\ldots
,B_M;\Sigma)=\sum_{s,t}\log \left[
  \sum_{k=1}^M \omega_k g\big(\boldsymbol x_{s,t}\vert \, A_k\!+\!B_k
  \boldsymbol f_{k,s,t},\Sigma \big) \right],
\end{equation}
where the first summation is over all locations \ $s\in{\mathcal S}$ \
and time points \ $t$ \ from the training period containing \ $N$ \
terms \ ($N$ \ distinct values of \ $(s,t)$).

To find the maximum of the log-likelihood \eqref{eq:eq3.6} we make use of the
EM algorithm for truncated normal mixtures suggested by \citet{ls12}.
Similarly to the traditional EM algorithm
for mixtures we introduce latent allocation variables
\ $z_{k,s,t}$ \ taking values one or zero according as whether \
$\boldsymbol x_{s,t}$ \ comes from the $k$th component PDF or not. The complete
data log-likelihood corresponding to the training data and allocations
equals
\begin{align*}
\ell_C (\omega_1,\ldots ,\omega_M;A_1, \ldots ,A_M; &\,B_1,\ldots
,B_M; \Sigma) \\
&=\sum_{s,t} \sum_{k=1}^M
z_{k,s,t} \bigg[\log(\omega_k) + \log \Big(
  g\big(\boldsymbol x_{s,t}\vert \, A_k+B_k \boldsymbol f_{k,s,t},
    \Sigma \big)\Big)\bigg].
\end{align*}

The EM algorithm starts with initial values  of
the parameters then alternates between an expectation (E) step and a
maximization (M) step until convergence. The coefficients of linear
regression of the validating observations on the corresponding ensemble members
can serve as initial values of \ $A_k^{(0)}$ \
and \ $B_k^{(0)}, \ k=1,2,\ldots ,M,$ \ the covariance matrix of the
validating observations can be taken as  \ $\Sigma^{(0)}$,  \ while the
initial weights \ $\omega_k^{(0)}, \ k=1,2,\ldots ,M,$ \ might be set
to be all equal.

For the truncated normal mixture model given by \eqref{eq:eq3.3} and
\eqref{eq:eq3.4} the E step is,
\begin{equation}
  \label{eq:eq3.7}
z_{k,s,t}^{(j+1)}:=\frac {\omega_k^{(j)}g\big(\boldsymbol x_{s,t}\vert \,
  A_k^{(j)}+B_k^{(j)} \boldsymbol f_{k,s,t}, \Sigma^{(j)} \big)}{\sum
  _{i=1}^M\omega_i^{(j)}g\big(\boldsymbol x_{s,t}\vert \,
  A^{(j)}_i+B^{(j)}_i \boldsymbol f_{i,s,t}, \Sigma^{(j)} \big)},
\end{equation}
where the superscript refers to the actual iteration. The first part
of the M step is
\begin{equation}
  \label{eq:eq3.8}
\omega_k^{(j+1)}:=\frac 1N\sum_{s,t}z_{k,s,t}^{(j+1)},
\end{equation}
while the second part can be derived from equations
\begin{equation}
  \label{eq:eq3.9}
\frac{\partial \ell_C}{\partial A_k} =0, \qquad \frac{\partial
  \ell_C}{\partial B_k} =0, \qquad \frac{\partial \ell_C}{\partial
  \Sigma} =0,  \qquad k=1,2, \ldots ,M.
\end{equation}
As the above system of equations is nonlinear, we suggest iteration steps
\begin{align}
A_k^{(j+1)}\!:=&\,\left[\sum_{s,t}
z_{k,s,t}^{(j+1)} \!\left(\Big (\boldsymbol x_{s,t}-
  B_k^{(j)}\boldsymbol f_{k,s,t}\Big)- \frac 1{\sigma_W^{(j)}}\frac{\varphi
  \Big(\mu_{W,k,s,t}^{(j)} /\sigma_W^{(j)} \Big)}{ \Phi
  \Big(\mu_{W,k,s,t}^{(j)}/\sigma_W^{(j)}\Big)}
  \begin{bmatrix}\big(\sigma_W^{(j)}\big)^2\\ \sigma_{WT}^{(j)}\end{bmatrix}
\right)\right]\bigg[\sum_{s,t}z_{k,s,t}^{(j+1)}\bigg]^{-1}\!\!\!,  \nonumber\\
B_k^{(j+1)}\!:=&\,\left[\sum_{s,t} z_{k,s,t}^{(j+1)} \!\left(\Big
    (\boldsymbol x_{s,t}- A_k^{(j+1)}\Big)-\frac 1{\sigma_W^{(j)}}\frac{\varphi
  \Big(\tilde\mu_{W,k,s,t}^{(j)} /\sigma_W^{(j)}  \Big)}{ \Phi
  \Big(\tilde\mu_{W,k,s,t}^{(j)} /\sigma_W^{(j)}\Big)}
  \begin{bmatrix}\big(\sigma_W^{(j)}\big)^2\\ \sigma_{WT}^{(j)}\end{bmatrix}
  \right)  \boldsymbol f_{k,s,t}^{\top}\right]  \nonumber\\
 &\times \bigg[\sum_{s,t}z_{k,s,t}^{(j+1)} \boldsymbol f_{k,s,t}
 \boldsymbol f_{k,s,t}^{\top}\bigg]^{-1}, \label{eq:eq3.10} \\
\Sigma^{(j+1)}\!:=&\,\frac 1N\sum_{s,t} \sum_{k=1}^Mz_{k,s,t}^{(j+1)} \!\left(\Big
    (\boldsymbol x_{s,t}- \boldsymbol\mu_
    {k,s,t}^{(j+1)}\Big)\Big(\boldsymbol x_{s,t} - \boldsymbol\mu_
    {k,s,t}^{(j+1)}\Big)^{\top} \right. \nonumber \\
    &\qquad \qquad \qquad \qquad \left. + \boldsymbol\mu_
    {k,s,t}^{(j+1)} \frac 1{\sigma_W^{(j)}}\frac{\varphi
  \Big(\mu_{W,k,s,t}^{(j+1)} /\sigma_W^{(j)}  \Big)}{ \Phi
  \Big(\mu_{W,k,s,t}^{(j+1)} /\sigma_W^{(j)}\Big)}
  \begin{bmatrix}\big(\sigma_W^{(j)}\big)^2& \sigma_{WT}^{(j)}\\
    \sigma_{WT}^{(j)}
    &\big(\sigma_{WT}^{(j)}/\sigma_W^{(j)}\big)^3\end{bmatrix}
\right), \nonumber
\end{align}
where \ $\mu_{W,k,s,t}^{(j)}$ \ and \ $\tilde \mu_{W,k,s,t}^{(j)}$ \ denote
the first (wind) coordinates of \
$\boldsymbol\mu_{k,s,t}^{(j)}:=A_k^{(j)}+B_k^{(j)}\boldsymbol f_{k,s,t}$ \ and \
\
$\tilde{\boldsymbol \mu}_{k,s,t}^{(j)}:=A_k^{(j+1)}+B_k^{(j)}\boldsymbol
f_{k,s,t}$,\  respectively.

\subsubsection*{Parsimonious model}
For the parsimonious model \eqref{eq:eq3.5} the log-likelihood
function is obviously
\begin{equation*}
\ell (\omega_1,\ldots ,\omega_M;A; B;\Sigma)=\sum_{s,t}\log \left[
  \sum_{k=1}^M \omega_k g\big(\boldsymbol x_{s,t}\vert \, A+B
  \boldsymbol f_{k,s,t},\Sigma \big) \right],
\end{equation*}
which is maximized using the same type of EM algorithm as before. The
E step, and the iterations corresponding to \ $\omega_k^{(j+1)}$ \ and \
$\Sigma^{(j+1)}$ \ are obvious modifications of \eqref{eq:eq3.7},
\eqref{eq:eq3.8} and of the last iteration of \eqref{eq:eq3.10},
respectively, while the first two iterations of \eqref{eq:eq3.10}
should be replaced by
\begin{align*}
A^{(j+1)}:=&\,\frac 1N\sum_{s,t} \sum_{k=1}^M
z_{k,s,t}^{(j+1)} \left(\Big (\boldsymbol x_{s,t}-
  B^{(j)}\boldsymbol f_{k,s,t}\Big)- \frac 1{\sigma_W^{(j)}}\frac{\varphi
  \Big(\mu_{W,k,s,t}^{(j)} /\sigma_W^{(j)} \Big)}{ \Phi
  \Big(\mu_{W,k,s,t}^{(j)}/\sigma_W^{(j)}\Big)}
  \begin{bmatrix}\big(\sigma_W^{(j)}\big)^2\\ \sigma_{WT}^{(j)}\end{bmatrix}
\right),  \nonumber\\
B^{(j+1)}:=&\,\sum_{s,t}\sum_{k=1}^M z_{k,s,t}^{(j+1)} \!\left(\Big
    (\boldsymbol x_{s,t}- A^{(j+1)}\Big)-\frac 1{\sigma_W^{(j)}}\frac{\varphi
  \Big(\tilde\mu_{W,k,s,t}^{(j)} /\sigma_W^{(j)}  \Big)}{ \Phi
  \Big(\tilde\mu_{W,k,s,t}^{(j)} /\sigma_W^{(j)}\Big)}
  \begin{bmatrix}\big(\sigma_W^{(j)}\big)^2\\ \sigma_{WT}^{(j)}\end{bmatrix}
  \right)  \boldsymbol f_{k,s,t}^{\top} \nonumber\\
 &\times \bigg[\sum_{s,t}\sum_{k=1}^Mz_{k,s,t}^{(j+1)} \boldsymbol f_{k,s,t}
 \boldsymbol f_{k,s,t}^{\top}\bigg]^{-1}.
\end{align*}
In this case \ $\mu_{W,k,s,t}^{(j)}$ \ and \ $\tilde \mu_{W,k,s,t}^{(j)}$ \ denote
the first coordinates of \
$\boldsymbol\mu_{k,s,t}^{(j)}:=A^{(j)}+B^{(j)}\boldsymbol f_{k,s,t}$ \ and \
\ $\tilde{\boldsymbol \mu}_{k,s,t}^{(j)}:=A^{(j+1)}+B^{(j)}\boldsymbol
f_{k,s,t}$,\  respectively.

\subsection{Multivariate scores}
  \label{subs:subs3.4}
To check the goodness of fit of bivariate probabilistic forecasts and of
the corresponding point forecasts we apply the methods suggested by
\citet{gsghj}.

For inspecting calibration of univariate ensemble forecasts a popular tool
is the verification rank histogram (or Talagrand diagram) which is the
histogram of ranks of validating observations with respect to the
corresponding ensemble forecasts \citep[see, e.g.,][Section
8.7.2]{wilks}. In case of proper calibration the ranks follow a uniform
distribution on \ $\{1,2,\ldots ,M+1\},$ \ and
the deviation from uniformity can be
quantified by the reliability index \ $\Delta$ \ defined by
\begin{equation}
   \label{eq:eq3.11}
 \Delta:=\sum_{r=1}^{M+1}\Big| \rho_r-\frac 1{M+1}\Big|,
\end{equation}
where \ $\rho_j$ \ is the relative frequency of rank \ $r$ \
\citep{delle}. In the multivariate case the usual problem is the
proper definition of ranks -- in the present work we use the
multivariate ordering proposed by \citet{gsghj}. For a probabilistic
forecast one
can calculate the reliability index from a preferably large number of
ensembles sampled from the predictive PDF and the corresponding
verifying observations.

In the univariate case, sharpness of an ensemble forecast or of a
predictive distribution can be
quantified by its standard deviation. An obvious generalization of
this idea to $d$-dimensional quantities is the determinant sharpness \
$\DS$ \ defined as
\begin{equation}
  \label{eq:eq3.12}
\DS:= \big(\det (\Sigma)\big)^{1/(2d)},
\end{equation}
where \ $\Sigma $ \ is the covariance matrix of an ensemble or of a
predictive PDF.

For evaluating density forecasts of univariate quantities the
continuous ranked
probability score (CRPS) is a widely
accepted and used proper scoring rule \citep{grjasa,wilks}.  A direct
multivariate extension of the CRPS is the energy score introduced by
\citet{grjasa}. Given a CDF \ $F$ \ on \
${\mathbb R}^d$ \ and a $d$-dimensional vector \ $\boldsymbol x$, \
the energy score is defined as
\begin{equation}
 \label{eq:eq3.13}
\ES(F,\boldsymbol x):={\mathsf E}\Vert \boldsymbol X-\boldsymbol
x\Vert-\frac 12 {\mathsf E}\Vert \boldsymbol X-\boldsymbol X'\Vert,
 \end{equation}
where \ $ \boldsymbol X$ \ and \  $\boldsymbol X'$ \ are independent
random vectors having distribution \ $F$. \ However, for a mixture of truncated
bivariate normal distributions considered in the present work the
energy score cannot be given in a closed form, so it is replaced by a
Monte Carlo approximation
\begin{equation}
  \label{eq:eq3.14}
\widehat\ES(F,\boldsymbol x):=\frac 1n \sum _{j=1}^n\Vert \boldsymbol
X_j-\boldsymbol x\Vert-\frac 1{2(n-1)}\sum_{j=1}^{n-1}\Vert
\boldsymbol X_j-\boldsymbol X_{j+1}\Vert,
 \end{equation}
where \ $\boldsymbol X_1,\boldsymbol X_2, \ldots ,\boldsymbol X_n$ \
is a (large, we use \ $n=10000$) \ random sample from \ $F$ \
\citep{gsghj}. Finally, if \
$F$ \ is a CDF corresponding to a forecast ensemble \ $\boldsymbol
f_1,\boldsymbol f_2, \ldots ,\boldsymbol f_M$ \ then \eqref{eq:eq3.13}
reduces to
\begin{equation}
  \label{eq:eq3.15}
\ES(F,\boldsymbol x)=\frac 1M \sum _{j=1}^M\Vert \boldsymbol
f_j-\boldsymbol x\Vert-\frac 1{2M^2}\sum_{j=1}^M\sum_{k=1}^M\Vert
\boldsymbol f_j-\boldsymbol f_k\Vert.
 \end{equation}

Finally, for point forecasts (mean and median) one can consider the
mean Euclidean
distance \ ($\EE$) \ of forecasts from the corresponding
validating observations. For multivariate forecasts the ensemble
median can be obtained using the Newton-type algorithm given in
\citet{ds83}, the algorithm of \citet{vz}, or any other method
implemented, e.g., in the R package  {\tt pcaPP}
\citep{ffc}. For a predictive distribution \ $F$ \ one may apply the
same algorithm on a preferably large sample from \ $F$.

\section{Results}
  \label{sec:sec4}
As it has been mentioned in the Introduction, the predictive
performances of the bivariate BMA models \eqref{eq:eq3.4} and
\eqref{eq:eq3.5} are tested on the eight-member UWME and
  on ensemble
forecasts produced by the ALADIN-HUNEPS system of the HMS.
We quantify the goodness of fit of the predictive PDFs and point
forecasts with the help of multivariate scores described in
Section \ref{subs:subs3.4}
and compare the results to the performance of the Gaussian copula method
proposed by \citet{mlt}.
For the case study conducted in \citet{mlt}, the univariate BMA
post-processing of the copula margins
is performed at each considered station individually, as the
performance of the method at specific stations as well as the
structure of correlations were investigated.
For the analysis in this paper the copula margins are formed by
applying a global BMA model to have a better comparability to the
bivariate truncated normal BMA model. This leads to the estimation of only
one single correlation matrix over all considered stations instead of
station specific correlation matrices.

\subsection{University of Washington mesoscale ensemble}
   \label{subs:subs4.1}

\subsubsection*{Raw ensemble}

\begin{figure}[t]
\begin{center}
\leavevmode
\epsfig{file=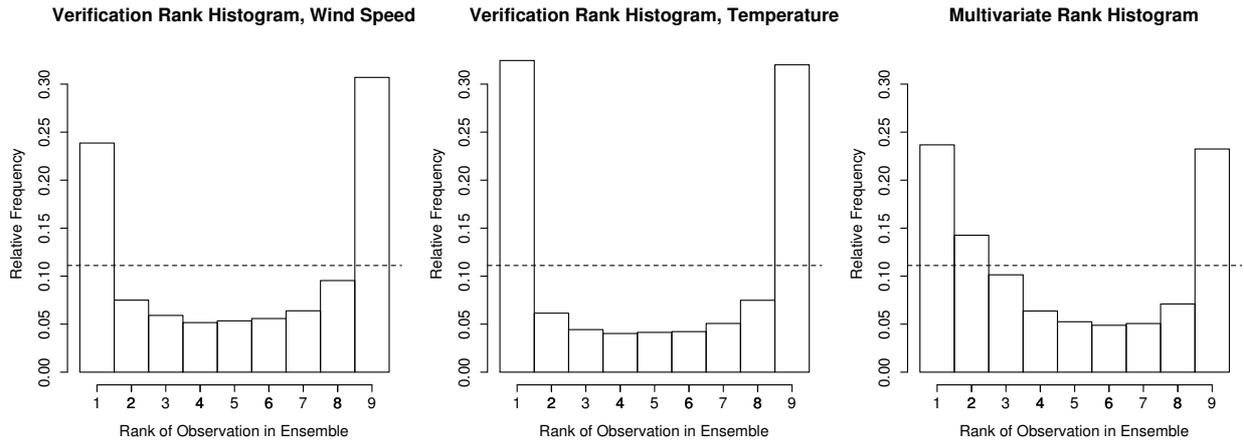,height=5.7cm, angle=0}
\caption{Verification rank histograms of the 8-member UMWE forecasts
  of maximum wind speed ({\em left}) and minimum temperature ({\em
    center}) and the multivariate rank histogram ({\em right}).
  Period: January 1,  2008 --  December 31, 2008.}
\label{fig:fig1}
\end{center}
\end{figure}

Earlier studies dealing with statistical calibration of the
University of Washington mesoscale ensemble \citep[see,
e.g.,][]{tg,frg} found that both wind speed and  
temperature forecasts are highly under-dispersive and in this way they are
uncalibrated. This underdispersive character can clearly be observed
in Figure \ref{fig:fig1} showing the univariate verification
histograms of wind speed and temperature as well as their joint
multivariate rank histogram. All three histograms are
far from the desired uniform distribution and in many cases the
ensemble members either underestimate or overestimate the validating
observations. The ensemble ranges contain the observed maximum wind
speed and minimum
temperature only in $45.44\% $ and  $35.53\%$ of the cases, respectively.
The reliability index $\Delta$ \ computed from the multivariate ranks
equals $0.5570$, while the $\Delta$ values corresponding to the
univariate ranks of wind speed and
temperature observations are given as $0.6468$ and $0.8449$,
respectively. Verifying observations of wind speed and temperature for
calendar year 2008 taken along all dates and
locations show a significant positive correlation of $0.1221$,
which justifies the need of a bivariate model for these weather
variables.

As the eight members of the UWME are non exchangeable, in what follows
we consider BMA models \eqref{eq:eq3.4} and \eqref{eq:eq3.5} with \ $M=8$.

\subsubsection*{Bivariate ensemble calibration}

In the present work we apply the same training period length of 40 days
as in \citet{mlt} which was determined with the help of an
exploratory data analysis on a subset of the data set. Since in our
rolling training periods for estimating the BMA parameters we can also
use data from calendar year 2007, BMA models can be produced for the whole
calendar year 2008.
This means 355 calendar days (after excluding dates with missing data)
and a total of $24832$ individual forecast cases.
Similar to the case study performed in \citet{mlt} involving the UWME
ensemble, the data from 2007 are utilized for correlation
estimation. The resulting global correlation matrix is
then employed for the analysis of the 2008 data. However,
  one should mention that as two of the 92 stations present in the
  final data described in Section \ref{subs:subs2.1} have no observations
  in 2007, they could not be employed for correlation estimation. To
  use the same stations in the BMA models that are utilized to
  estimate the correlation matrix as in the BMA models employed for
  prediction, these two stations (meaning 530 forecast cases) are
  deleted for the copula method. 

\begin{table}[t!]
\begin{center}{\footnotesize
\begin{tabular}{lccccccc}
&$\ES$&$\Delta$&$\DS$&$\EE$ median&$\EE$ mean&$\varrho$
median&$\varrho$ mean\\ \hline
BMA&$2.1124$&$0.0155$&$2.5678$&$2.9768$&$2.9764$&$0.1455$&$0.1476 $\\
BMA
pars.&$2.1189$&$0.0330$&$3.2926$&$2.9710$&$2.9722$&$0.1733$&$0.1677$\\
\hline
Copula&$2.0894$&$0.0298$&$2.2723$&$2.9771$&$2.9817$&$0.1602$&$0.1824$\\ \hline
Raw ensemble&$2.5660$&$0.5570$&$0.7736$&$3.0916$&$3.0765$&$0.0162$&$0.0057$

\end{tabular}
\caption{Mean energy score ($\ES$), reliability index
  ($\Delta$) and mean determinant sharpness ($\DS$) of probabilistic
  forecasts, mean Euclidean  error ($\EE$) of point forecasts (bivariate
  median/mean) and empirical correlation ($\varrho$) of wind speed and
  temperature  components of point forecasts for the
  UWME.} \label{tab:tab1} }
\end{center}
\end{table}

\begin{figure}[t]
\begin{center}
\leavevmode
\epsfig{file=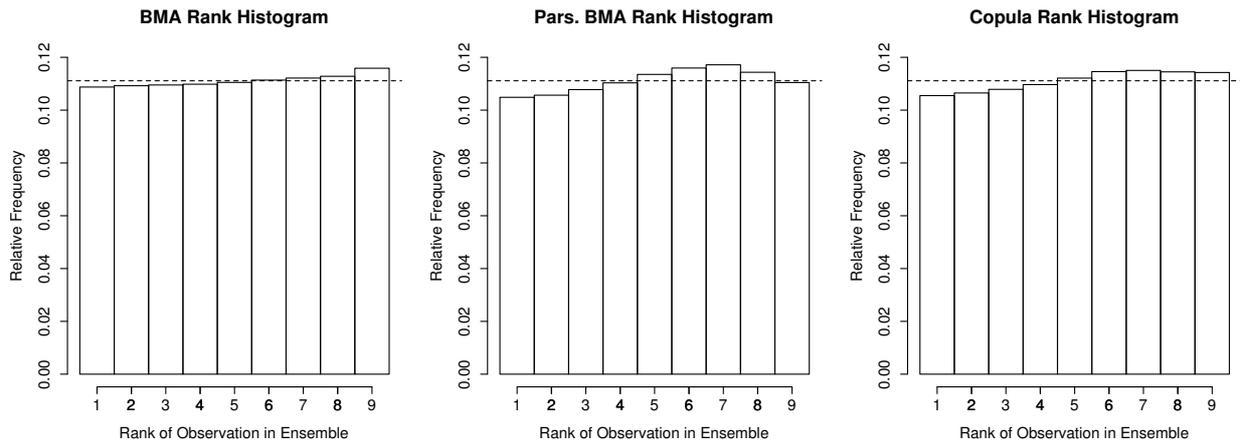,height=5.7cm, angle=0}
\caption{Multivariate rank histograms for BMA ({\em left}),
  parsimonious BMA ({\em center}) and Gaussian copula ({\em right})
  post-processed UWME forecasts of maximum wind speed and minimum temperature.}
\label{fig:fig2}
\end{center}
\end{figure}

Table \ref{tab:tab1} shows the mean energy score ($\ES$), reliability
index ($\Delta$)
and mean determinant sharpness ($\DS$) of probabilistic forecasts and the
mean Euclidean error ($\EE$) of point forecasts together with empirical
correlation of their wind speed and temperature components calculated
using both bivariate BMA models considered, the copula model of
\citet{mlt} and the raw ensemble. All three post-processing
methods result in significant improvement in calibration of the
probabilistic forecasts, quantified by the decrease of
the mean energy score and reliability index,
and in a slight improvement in accuracy of median and mean
forecasts (see the corresponding \ $\EE$ \ values). The
improvement in calibration can clearly be observed on the difference
between the U-shaped multivariate rank histogram of the raw ensemble (see
Figure \ref{fig:fig1}) and the
rank histograms of post-processed forecasts plotted in Figure
\ref{fig:fig2}, which are almost uniform.
Furthermore,
the empirical correlations of wind speed and temperature components of
all post-processed point forecasts are close to the
correlation of $0.1221$ of the verifying wind speed and temperature
observations, while the corresponding correlations of ensemble median
and mean are smaller by a magnitude. The \ $\DS$ \
of the predictive PDFs is much higher than that of the raw ensemble, however,
this is a direct consequence of the small dispersion of the latter
(see Figure \ref{fig:fig1}).
Comparing the different post-processing methods it is noticeable that
there is no big difference between their predictive performances.
The bivariate truncated normal BMA models produce slightly more
accurate point forecasts, while the smallest
\ $\ES$ \ and \ $\DS$ \ values correspond to the copula method. The
general BMA model \eqref{eq:eq3.4} outperforms its parsimonious
counterpart \eqref{eq:eq3.5} and from the three competing methods this
results in the best reliability index. This is in line with the shapes
of the corresponding multivariate rank histograms in Figure
\ref{fig:fig2}. The rank histogram of the general BMA model is closest
to uniformity, while the rank histograms of the parsimonious BMA and
the copula model both exhibit similar deviations from the uniform distribution. 

\subsection{ALADIN-HUNEPS ensemble}
   \label{subs:subs4.2}

\subsubsection*{Raw ensemble}
\begin{figure}[t]
\begin{center}
\leavevmode
\epsfig{file=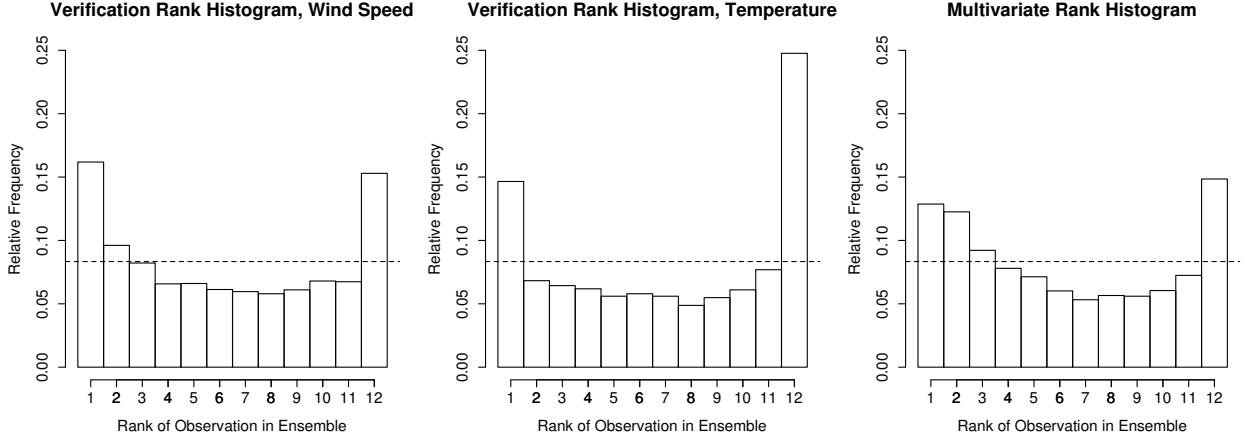,height=5.7cm, angle=0}
\caption{Verification rank histograms of the 11-member ALADIN-HUNEPS
  ensemble forecasts of  wind speed ({\em left}) and temperature ({\em
    center}) and the multivariate rank histogram ({\em right}).
  Period: April 1,  2012 --  March 31, 2013.}
\label{fig:fig3}
\end{center}
\end{figure}

Similar to the UWME, wind speed and temperature
forecasts of the ALADIN-HUNEPS ensemble are found to be underdispersive
\citep{bhn1,bhn2}. The univariate verification rank
histograms of wind speed and temperature as well as their joint
multivariate rank histogram plotted in Figure \ref{fig:fig3} are
strongly U-shaped and
the ensemble ranges contain the observed wind speed and
temperature only in $68.52\% $ and  $60.61\%$ of the cases, respectively.
The reliability index computed from the multivariate ranks equals
$0.3172$, while the reliability indices obtained from the univariate
ranks of wind speed and
temperature observations are $0.3217$ and $0.4550$, respectively.
Observations of wind speed and temperature taken along all dates and
locations show a slight negative
correlation of $-0.0294$, which is not significant under the
assumption of independence in space and time. This difference compared
to the UWME might be explained by the different types of wind  and
temperature quantities being analyzed.

Using the ideas of \citet{bhn1,bhn2} we consider two different
groupings of ensemble members. In the first case we have two
exchangeable groups. One contains the control denoted by \
$\boldsymbol f_c$ \ while the other group contains the remaining 10
ensemble members
corresponding to the different perturbed initial conditions denoted by
\ $\boldsymbol f_{p,1},\ldots ,\boldsymbol f_{p,10}$. \  This
leads us to the predictive PDF
\begin{align}
  \label{eq:eq4.1}
p_{AH2}\big(\boldsymbol x |\, \boldsymbol f_c\,\boldsymbol f_{p,1},\ldots ,
\boldsymbol f_{p,10};A_c,A_p;B_c,B_p;\Sigma\big) = &\,\omega
g\big(\boldsymbol x|\,A_c+B_c \boldsymbol f_c, \Sigma\big) \\
&+\frac {1-\omega}{10} \sum_{\ell=1}^{10}
g\big(\boldsymbol x|\,A_p+B_p\boldsymbol f_{p,\ell},\Sigma\big),  \nonumber
\end{align}
which is a particular case of model \eqref{eq:eq3.4}, and to its
parsimonious version
\begin{equation}
  \label{eq:eq4.2}
q_{AH2}\big(\boldsymbol x |\, \boldsymbol f_c\,\boldsymbol f_{p,1},\ldots ,
\boldsymbol f_{p,10};A_;B;\Sigma\big) = \,\omega
g\big(\boldsymbol x|\,A+B \boldsymbol f_c, \Sigma\big)
+\frac {1-\omega}{10} \sum_{\ell=1}^{10}
g\big(\boldsymbol x|\,A+B\boldsymbol f_{p,\ell},\Sigma\big)
\end{equation}
corresponding to model \eqref{eq:eq3.5}, where \ $\omega\in [0,1]$, \ and \
$g$ \ is defined by
\eqref{eq:eq3.3}.

In the second case the odd and even numbered exchangeable ensemble
members form two separate groups
$\{\boldsymbol f_{p,1}, \ \boldsymbol f_{p,3}, \ \boldsymbol f_{p,5},
\ \boldsymbol f_{p,7}, \ \boldsymbol f_{p,9}\}$ \ and \ $\{\boldsymbol
f_{p,2}, \ \boldsymbol f_{p,4}, \ \boldsymbol f_{p,6}, \ \boldsymbol
f_{p,8}, \ \boldsymbol f_{p,10}\}$, \ respectively. This idea is
justified by the method of generating their initial conditions.
To obtain the initial conditions for the ALADIN-HUNEPS forecasts
only five perturbations are calculated and
then they are added to (odd numbered members) and
subtracted from (even numbered members) the unperturbed
initial conditions \citep{horanyi,bhn1,bhn2}. In
this way we obtain the following PDFs for the forecasted vector of wind
speed and temperature corresponding to models \eqref{eq:eq3.4} and
\eqref{eq:eq3.5}, respectively,
\begin{align}
  \label{eq:eq4.3}
p_{AH3}\big(\boldsymbol x |\, \boldsymbol f_c,\boldsymbol f_{p,1},\ldots ,
\boldsymbol f_{p,10};&\,A_c, A_o, A_e; B_c,B_o,B_e;\Sigma
\big)= \omega_c
g\big(\boldsymbol x|\,A_c+\boldsymbol f_cB_c,\Sigma\big) \\ &+
\sum_{\ell=1}^{5} \Big(\omega_o g\big(\boldsymbol
x|\,A_o+B_o\boldsymbol f_{p,2\ell -1},\Sigma\big)+
\omega_e g\big(\boldsymbol x|\,A_e+B_e\boldsymbol
f_{p,2\ell},\Sigma\big)\Big), \nonumber 
\end{align}
\begin{align}
q_{AH3}\big(\boldsymbol x |\, \boldsymbol f_c,\boldsymbol f_{p,1},\ldots ,
\boldsymbol f_{p,10};&\,A; B;\Sigma \big)= \omega_c
g\big(\boldsymbol x|\,A+\boldsymbol f_cB,\Sigma\big)  \label{eq:eq4.4}
\\ &+
\sum_{\ell=1}^{5} \Big(\omega_o g\big(\boldsymbol
x|\,A+B\boldsymbol f_{p,2\ell -1},\Sigma\big)+
\omega_e g\big(\boldsymbol x|\,A+B\boldsymbol
f_{p,2\ell},\Sigma\big)\Big), \nonumber
\end{align}
where for weights \ $\omega_c,\omega_o,\omega_e\in[0,1]$ \ we have  \
$\omega_c+5\omega_o+5\omega_e=1$.

\subsubsection*{Bivariate ensemble calibration}

\begin{table}[t!]
\begin{center}{\footnotesize
\begin{tabular}{lccccccc}
&$\ES$&$\Delta$&$\DS$&$\EE$ median&$\EE$ mean&$\varrho$
median&$\varrho$ mean\\ \hline
Two-group BMA&$1.4344$&$0.0310$ 
&$1.8544$&$2.0043$&$2.0063$&$-0.0322$&$-0.0393$\\
Three-group
BMA&$1.4423$&$0.0308$&$1.7250$&$2.0167$&$2.0199$&$-0.0345$&$-0.0375$\\
\hline
Two-group BMA
pars.&$1.4280$&$0.0209$&$1.9285$&$1.9989$&$1.9974$&$-0.0311$&$-0.0338$\\
Three-group BMA
pars.&$1.4283$&$0.0207$&$1.9236$&$1.9976$&$1.9965$&$-0.0301$&$-0.0330$\\
\hline
Two-group copula&$1.3929$&$0.0630$&$1.5258$&$2.0319$&$2.0402$&$-0.0205$&$0.0166$\\
Three-group copula&$1.3886$&$0.0656$&$1.5205$&$2.0291$&$2.0365$&$-0.0188$&$0.0180$\\ \hline
Raw ensemble&$1.6232$&$0.3270$&$0.9352$&$2.1019$&$2.0830$&$-0.0684$&$-0.0603$

\end{tabular} }
\caption{Mean energy score ($\ES$), reliability index
  ($\Delta$) and mean determinant sharpness ($\DS$) of probabilistic
  forecasts, mean Euclidean  error ($\EE$) of point forecasts (bivariate
  median/mean) and empirical correlation ($\varrho$) of wind speed and
  temperature  components of point forecasts for the  ALADIN-HUNEPS
  ensemble.} \label{tab:tab2}
\end{center}
\end{table}

Based on a preliminary data analysis (univariate BMA and EMOS
calibration of wind speed and temperature forecasts) we use a 40 days
training period. In this way ensemble members, validating observations
and BMA models are available for the period 12.05.2012--31.03.2013
(just after the first 40 days training period having 318 calendar
days, since on six days all ensemble members are missing).
In line with the case study performed in \citet{mlt}, additional data
of the period 01.10.2010--25.03.2011 are utilized to
estimate the correlation matrix of the Gaussian copula. For the BMA
fits that are employed to estimate this correlation structure, a 40
days training period
was used as well. The resulting (global) correlation
matrix is then carried forward into the analysis of the 2012/2013
data.

\begin{figure}[t]
\begin{center}
\leavevmode
\epsfig{file=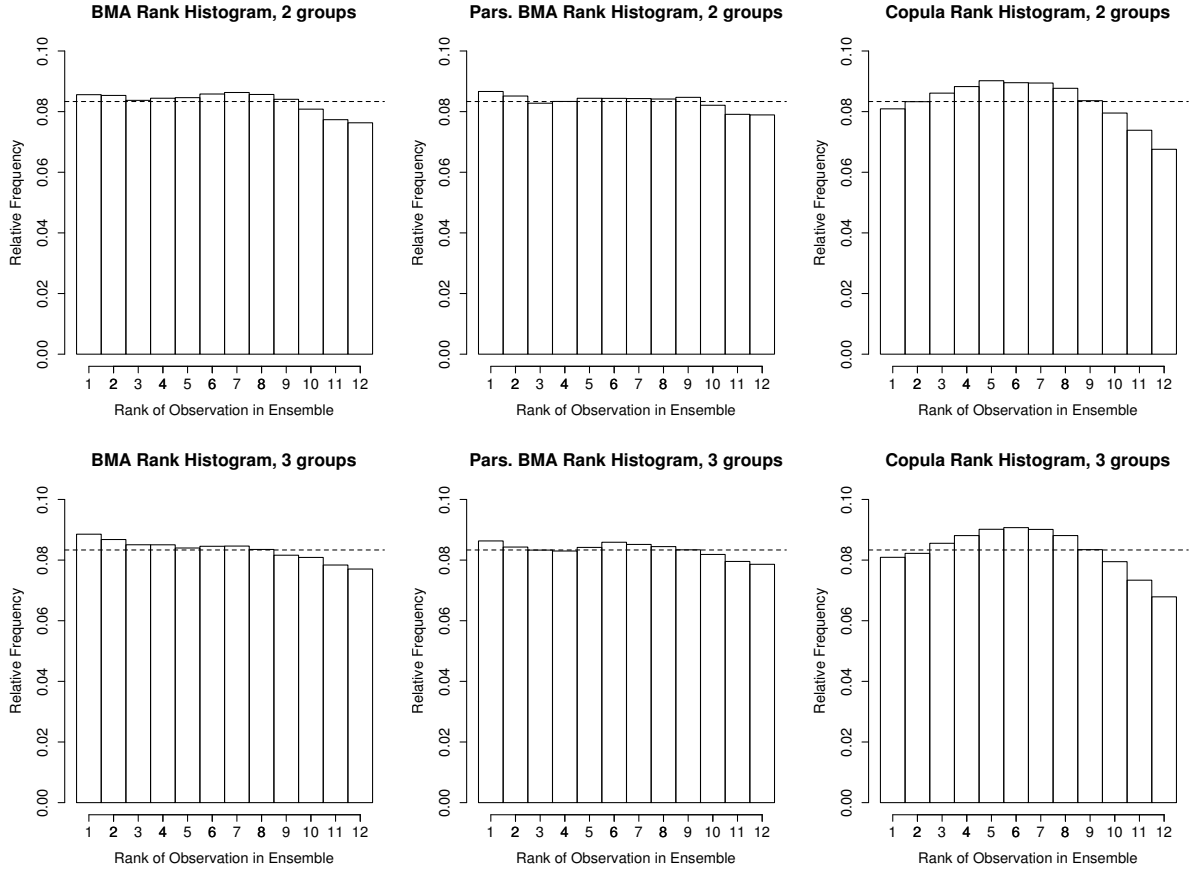,height=11.4cm, angle=0}
\caption{Multivariate rank histograms for BMA ({\em left}),
  parsimonious BMA ({\em center}) and Gaussian copula ({\em right})
  post-processed ALADIN-HUNEPS forecasts of wind speed and temperature
  using two- and three-group models.}
\label{fig:fig4}
\end{center}
\end{figure}

The verification scores quantifying the effects of ensemble post-processing
are given in Table \ref{tab:tab2}. Considering first the probabilistic
forecasts one can observe that compared to the raw ensemble the BMA
and copula predictive PDFs are smaller in
energy score and reliability index and higher in determinant
sharpness. Again, the latter fact comes from the underdispersive
character of the raw ensemble illustrated by Figure
\ref{fig:fig3}, while the rank histograms of the post-processed forecasts,
plotted in Figure \ref{fig:fig4}, clearly illustrate the improvement in
calibration. Regarding median/mean forecasts calculated from
the above mentioned predictive PDFs, they all produce smaller
\ $\EE$ \ values than the ensemble median/mean vectors. Furthermore,
the wind speed and temperature components of the
ensemble median/mean show a significant negative correlation, while
the correlations corresponding to the post-processed point forecasts
follow the correlation of the validating observations of the two
weather quantities. In contrast to the results for the UWME, here
one can find a significant difference between the bivariate BMA and copula
approaches. The bivariate BMA post-processing results in much smaller
\ $\Delta$ \ values than the corresponding reliability indices of the
copula method. This is in line with the multivariate rank histograms in 
Figure \ref{fig:fig4}. The histograms of both bivariate BMA methods
are closer to uniformity than the one of the copula method, for the
two- as well as for the three-group model.  
The histogram of the copula method exhibits a slight forecast bias,
the last bins are less filled than the others.  
This phenomenon is also present in the histograms of the bivariate BMA
models, but much less pronounced.  
Further, the bivariate BMA post-processing model yields
slightly smaller Euclidean errors, while the copula
model produces sharper predictive PDFs and a bit smaller mean
energy scores. In general, three-group models result in better
calibrated and more accurate forecasts than their
two-group counterparts and the parsimonious BMA models \eqref{eq:eq4.3}
and \eqref{eq:eq4.4} outperform (in terms of $\ES, \ \Delta$ and
$\EE$) the general bivariate
BMA models \eqref{eq:eq4.1} and \eqref{eq:eq4.2}, respectively.

\section{Discussion}
  \label{sec:sec5}

In the present study we introduce a new bivariate BMA model for joint
calibration of ensemble forecasts of wind speed and temperature
providing a predictive PDF which is a mixture of bivariate normal
distributions truncated from below at zero in their first (wind)
coordinates. Two approaches are presented: a general and a
parsimonious one, differing only in the number of parameters to be
estimated. The models are tested on the eight-member UWME and on the
eleven-member ALADIN-HUNEPS ensemble of the HMS. The two ensemble
prediction systems differ both in generation of ensemble members and in
wind speed and temperature quantities being forecasted. The predictive
performances of both BMA post-processing methods,
quantified by the energy score, reliability index and determinant
sharpness of probabilistic and Euclidean error of point forecasts
(median and mean), are compared to the performance of the Gaussian
copula method suggested by \citet{mlt}.

In case of the UWME forecast vectors of maximum wind speed and
minimum temperature a 40 days rolling training period is used. Compared to
the raw ensemble all three post-processing methods significantly improve the
calibration of probabilistic and accuracy of point forecasts. Further,
the correlations of the calibrated point forecasts of wind speed and
temperature are very close to the empirical
correlation of the validating observations of these weather
quantities, while the components of the ensemble mean and median
vectors are practically uncorrelated. There is no big difference
between the verification scores of the two bivariate BMA models and of the
copula model: the latter results in the smallest mean energy score and
determinant sharpness, the smallest reliability index belongs to the
general bivariate BMA method, while the parsimonious bivariate BMA
model produces the smallest Euclidean errors among the point forecasts.

For calibrating ensemble forecast vectors of instantaneous wind speed
and temperature produced by the ALADIN-HUNEPS system a training period
of length 40 days and two different grouping of exchangeable ensemble
members are considered: one assumes two groups (control
and forecasts from perturbed initial conditions), while the other considers
three (control and forecasts from perturbed initial conditions
with positive and negative perturbations). According to the
verification scores investigated, the overall performances of
three-group models are slightly better than those of their two-group
counterparts, which is in accordance with the results of univariate
BMA calibration. The comparison of
the raw ensemble and of the post-processed forecast again shows the
positive effect of calibration resulting in smaller \ $\ES, \ \Delta$
\ and \ $\EE$ \ values. Moreover, the components of the ensemble mean and median
vectors show a significant negative correlation, while the small
correlations of the post-processed wind speed and temperature
forecasts give back the lack of correlation of the verifying
observations. Compared to the copula model of \citet{mlt} both BMA
approaches yield slightly lower Euclidean errors, slightly higher
mean energy scores, higher \ $\DS$ \ values and significantly lower
reliability indices. From the three competing post-processing methods
the overall performance of the parsimonious bivariate BMA model seems to be the
best.

It should be noted, that for both considered forecast ensembles, the
copula method yields a higher level of sharpness than the bivariate
BMA models, which, on the contrary exhibit a higher level of
calibration, indicated both by the corresponding $\Delta$ values and
multivariate rank histograms.  
This may be explained by the different covariance
structures of the methods. The slightly higher spread in the
bivariate BMA predictive distribution has a positive effect on the level of
calibration. Especially in case of the ALADIN-HUNEPS ensemble, 
the multivariate rank histogram for copula method still exhibit a
slight forecast bias, that is much less pronounced 
for the bivariate BMA model.

Therefore, we conclude that joint BMA post-processing of ensemble
predictions of wind speed and temperature significantly improves
calibration and accuracy of forecasts and the predictive performance
of the bivariate BMA model is at least as good as the performance of
the Gaussian copula approach of \citet{mlt}.

Finally, one should remark that the Gaussian copula method can be
applied for any desired type and number of weather quantities, while
the current version of the bivariate BMA model is applicable only for
a bivariate weather quantity vector where the components can be
assumed to be normal and truncated normal. However, an extension to a
higher dimensional setting is possible, although it might be computationally
challenging. As the ECC methodology proposed by \citet{stg13} is even
more flexible (and computationally more efficient) than the copula
method, a comparison of the bivariate BMA to ECC on exchangeable
ensemble forecasts (such as the ECMWF ensemble) might yield further
improvement of bivariate ensemble calibration.

\bigskip
\noindent
{\bf Acknowledgments.} \  \
Essential part of this work was made during the visiting professorship
of S\'andor Baran at the
Institute of Applied Mathematics of the University of Heidelberg.
Annette M\"oller gratefully acknowledges support by the German
Research Foundation (DFG)
within the program
``Spatio-/Temporal Graphical Models and Applications in Image
Analysis'' grant RTG 1653 in Heidelberg.
S\'andor Baran was supported by
the Hungarian  Scientific Research Fund under Grant No. OTKA NK101680
and by the T\'AMOP-4.2.2.C-11/1/KONV-2012-0001
project. The project has been supported by the European Union,
co-financed by the European Social Fund.
The authors are indebted to Tilmann
Gneiting and Andr\'as Hor\'anyi for their useful suggestions and
remarks, to the University of Washington MURI group for providing the
UWME data, to Mih\'aly Sz\H ucs from the HMS for the
ALADIN-HUNEPS data and to Thordis Thorarinsdottir and Alex Lenkoski for their
help with the R codes for the copula method.

\end{document}